# Effect of Multiple Scattering on the Critical Electric Field for Runaway Electrons in the Atmosphere

Ya. S. Elensky

*Institute for Nuclear Research, Russian Academy of Sciences, pr. Shestidesyatiletiya Oktyabrya 7a, Moscow, 117312 Russia*
*e-mail: yansa@gcnet.ru*

A simple method for taking into account the multiple Coulomb scattering in construction of a separatrix (the line separating the regions of runaway and decelerating electrons in an electric field) is described. The desired line is obtained by solving a simple transcendental equation.

DOI: 10.3103/S1062873807070428

## INTRODUCTION

The so-called separatrix—a line separating the runaway and deceleration regions is reported in almost each study devoted to runaway electrons. In the simplest case, the kinetic electron energy $T$ is plotted on the horizontal axis and the energy losses or the corresponding strength of the accelerating field ($F_{loss} = dT/dx$ or $F_{el} = dU/dx$) is plotted on the vertical axis. In this case, the separatrix is a dependence of the loss on energy; the point corresponding to a given electron energy $T$ and a given field $E_{el}$ lies either above or below the curve. In the first case, the electron is accelerated (runs away), and, in the second case, it is decelerated. Separatrices are plotted in such a form in [1–3]. In construction of these curves, multiple Coulomb scattering was disregarded. Nevertheless, it is obvious that a scattered electron changes the motion direction with respect to the field and, when the angle between the motion direction and field increases, a particle is weaker accelerated and stronger (due to the increase in the path length) decelerated. It seems that the curve should shift upwards; however, the situation is complicated by the fact that scattering is a random process, and, therefore, it is not obvious if the particle is accelerated or decelerated. We can only speak about the probability of being accelerated or decelerated. Let us consider the separatrix to be a curve connecting the points characterized by the same escape probability. In this statement of the problem, the "old" separatrix connects the points with zero escape probability. The purpose of the calculations reported below is to construct a separatrix for the case where most beam electrons are accelerated (the escape probability is close to unity).

## STATEMENT OF THE PROBLEM

Let us consider a beam of $N$ electrons propagating downward, along the accelerating field directed parallel to the $z$ axis (also directed downward). The beam can be described by the distribution function $n(p)$ in the space of momenta $p = \mathbf{p}(z)$ or $p = \mathbf{p}(t)$; here, $n$ is the number of particles per volume unit $dp$, and $\int n(\mathbf{p})d\mathbf{p} = N$ (one particle with the probability density function $f(p)$ can be also considered; in this case, $f(p) = n(p)/N$ and $\int f(\mathbf{p})d\mathbf{p} = 1$).

We can easily determine the fraction of runaway electrons if the function $n(p)$ is known. To this end, we have to solve the kinetic equation with appropriate initial conditions. However, according to [4], it is impossible to determine $n(p)$ analytically, and an adequate approximation should be used.

Numerical calculations were reported in [1–3], where the solution is sought for particular cases of electron motion. These calculations are fairly complicated and, apparently, not very reliable since they yield different results. In this study, we also perform numerical calculations, bearing in mind their maximum simplification with conservation of reasonable accuracy.

## CONSIDERATION OF SCATTERING

Analysis shows that a significant (if not the main) source of uncertainty is the incorrect consideration of electron scattering in air. For example, the ionization loss function was used in [2, 5] to calculate the scattering; this approach is justified for electron–electron scattering but has a poor accuracy for electron scattering by nuclei (the case where energy losses are almost absent). The difference is especially high at low energies (<100 keV).

A formula for the root-mean-square scattering angle after passage of $X$ radiation units is given in [6, 7]:

$$\sqrt{\langle \theta^2 \rangle} = E_S \frac{\sqrt{X}}{p\beta}. \qquad (1)$$

Here, $p$ is the momentum of a particle, $\beta$ is its velocity, and $E_S = 21$ MeV is a characteristic energy of multiple scattering. This formula is exact only for fast particles ($E > 100$ MeV); for slow particles, it gives a very large maximum angle (several rad) for each scattering and, thus, the result is overestimated.

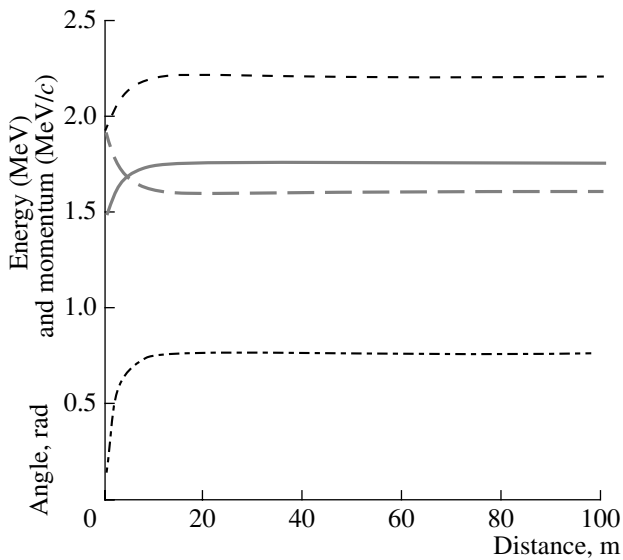

**Fig. 1.** Solution to the equation for $D = 3$ kV cm$^{-1}$ and $T_0 = 1.4875$ MeV: kinetic energy (solid line), momentum (short-dashed line), projection of the momentum on the field direction (long-dashed line), and the angle between the momentum and field (dash-dotted line).

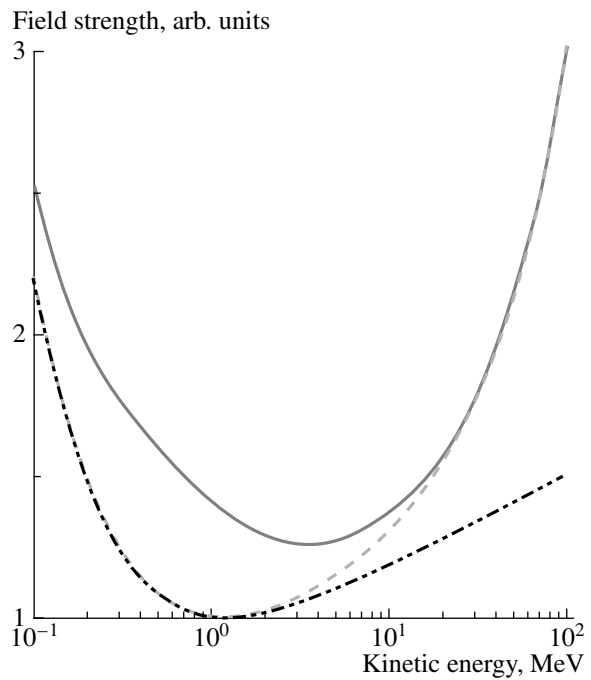

**Fig. 2.** Separatrix plotted with consideration of electron scattering (solid line), ionization and bremsstrahlung (dashed line), and only ionization (dash-dotted line).

Somewhat other formula was reported in [8]:

$$v\frac{d\langle\theta^2\rangle}{ds} = 2\pi N_a \left(\frac{2Z_a r_0 c^2}{\gamma v^2}\right)^2 \ln\left(\frac{\theta_{max}}{\theta_{min}}\right), \quad (2)$$

where $s$ is the passed distance, $r_0$ is the classical electron radius, and $v$ is the electron velocity. The expression under the logarithm is transformed in [8] as follows:

$$130.60 p/Z_a^{1/3} mc; \quad (2a)$$

however, the numerical coefficient may range from 65.3 to 384. The above examples show that the accuracy of solution of the kinetic equation is low because changes in angles are directly related to the coefficients of this equation. Hence, it is reasonable to simplify it, as, for example, in [8].

We will consider not the entire ensemble but only an "average" particle, for which

$$\theta = \sqrt{\langle\theta_{ens}^2\rangle}, \quad |p| = \langle|p_{ens}|\rangle.$$

Then, it is reasonable to replace the kinetic equation with the equation of motion

$$v\frac{d\mathbf{p}}{ds} = e\mathbf{D} + \mathbf{F}, \quad (3)$$

where $eD$ is the electric field force; $F = F_{loss} + F_{scat}$ is the sum of medium reactions; $F_{loss}$ is a force antiparallel to the momentum, which corresponds to losses; and $F_{scat}$ is a force directed perpendicular to the momentum, which arises due to scattering. Further simplification is possible if we take into account the properties of normal distribution. Indeed, after several (10–20) scatterings, the polar angle distribution is described by the normal circular law [6, 7], and the beam in space of momenta is a set of vectors with certain $|p|$ and $\theta$. The values of $|p|$ and $\theta$ change according to the equation of motion at a transition to some new level at a larger depth; however, the distribution remains normal. In this case, an arbitrary particle is randomly scattered so that the force $F_{scat}$ is uniformly distributed over the azimuthal angle. To conserve the particle in the "average" state, the force should be directed over radius from the beam axis to the periphery; however, in this case, it is not random any more. This consideration makes it possible to significantly simplify the solution of Eq. (3).

To solve this equation, the medium (air) is divided into thin horizontal layers $\Delta z = \cos(\theta)\Delta s$; in each such layer, the right-hand side of the equation is approximately constant, which makes it possible to calculate $\Delta p$. The calculation error due to the discreteness is smaller than 5%, as can be seen from the comparison of the results of calculations with different steps (1 and 10 cm). The main source of error is the deviation of the distribution of $\theta$ from the normal one due to the existence of a long "tail." Being not absolutely exact, the calculation makes it possible to see the characteristic features of motion. For example, the solution for an electron that is neither runaway nor decelerated (is on the separatrix) is shown in Fig. 1. The initial conditions are $\theta = 0$ and $T_{kin} = 1.4535$ MeV ($D = 3$ kV cm$^{-1}$). First, the angle $\theta$ increases until reaches the critical value, at which the deceleration and scattering are completely balanced by the electric field effect. Analysis shows that each value of the kinetic energy corresponds to a certain field strength and critical angle (plateau in Fig. 1). Note that the critical angles may be relatively large (about unity); hence, there are no limitations characteristic of the small-angle approximation.

## CONCLUSIONS

To construct the separatrix, we will take into account that in the case of force equilibrium the left-hand side of Eq. (3) becomes zero and we have the following transcendental equation:

$$0 = e\mathbf{D} + \mathbf{F}. \tag{3a}$$

Solving this equation for different energies, we obtain the desired separatrix. The results are shown in Fig. 2.

## REFERENCES

1. Gurevich, A.V. and Zybin, K.P., *Usp. Fiz. Nauk*, 2001, vol. 171, no. 11, p. 1177.
2. Gurevich, A.V., Milikh, G.M., and Roussel-Dupre, R.A., *Phys. Lett. A*, 1994, vol. 187, p. 197.
3. Lehtinen, N.G., Bell, T.F., and Inan, U.S., *J. Geophys. Res.*, 1999, vol. 104, p. 2469.
4. Remizovich, V.S., Rogozkin, D.B., and Ryazanov, M.I., *Fluktuatsii probegov zaryazhennykh chastits* (Fluctuations of Charged Particle Ranges), Moscow: Energoatomizdat, 1988.
5. Gurevich, A.V., Milikh, G.M., and Roussel-Dupre, R.A., *Phys. Lett. A*, 1992, vol. 165, p. 463.
6. Hayakawa, S., *Cosmic-Ray Physics*, New York: Wiley, 1969. Translated under the title *Fizika kosmicheskikh luchei*, Moscow: Mir, 1973.
7. Rossi, B. and Greizen, K., *Vzaimodeistvie kosmicheskikh luchei s veshchestvom* (Interaction of Cosmic Rays with Matter), Moscow: GIIL, 1948.
8. Lehtinen, N.G., PhD Thesis, Stanford: Stanford University, 2000